\documentclass[aps,prl,preprint,nopacs,superscriptaddress]{revtex4}
\usepackage{amsmath}
\usepackage{amssymb}
\usepackage{graphicx}
\usepackage{hyperref}
\usepackage[utf8]{inputenc}
\newcommand{\beq}{\begin{equation*}}
\newcommand{\eeq}{\end{equation*}}
\newcommand{\s}{Co$_3$Sn$_2$S$_2$}
\newcommand{\f}{$E_\textrm{F}$}

\newcommand{\eb}{$E_{\rm B}$}

\newcommand{\tc}{$T_{\rm C}$}
\newcommand{\ab}{{\it ab initio}}
\newcommand{\Ab}{{\it Ab initio}}

\newcommand{\pana}{a}
\newcommand{\panb}{b}
\newcommand{\panc}{c}
\newcommand{\pand}{d}
\newcommand{\pane}{e}
\newcommand{\panf}{f}
\newcommand{\pang}{g}
\newcommand{\panh}{h}

\newcommand{\cpana}{({\bf a})}
\newcommand{\cpanb}{({\bf b})}
\newcommand{\cpanc}{({\bf c})}
\newcommand{\cpand}{({\bf d})}
\newcommand{\cpane}{({\bf e})}
\newcommand{\cpanf}{({\bf f})}
\newcommand{\cpang}{({\bf g})}
\newcommand{\cpanh}{({\bf h})}

\begin{document}

\title{Signatures of Weyl fermion annihilation\\ in a correlated kagome magnet}

\author{Ilya Belopolski\footnote{These authors contributed equally to this work.}} \email{ilya.belopolski@riken.jp}
\affiliation{Laboratory for Topological Quantum Matter and Spectroscopy (B7), Department of Physics, Princeton University, Princeton, New Jersey 08544, USA}
\affiliation{RIKEN Center for Emergent Matter Science (CEMS), Wako, Saitama 351-0198, Japan}

\author{Tyler A. Cochran$^*$}
\affiliation{Laboratory for Topological Quantum Matter and Spectroscopy (B7), Department of Physics, Princeton University, Princeton, New Jersey 08544, USA}

\author{Xiaoxiong Liu$^*$}
\affiliation{Department of Physics, University of Zurich, Winterthurerstrasse 190, 8057 Zurich, Switzerland}

\author{Zi-Jia Cheng$^*$}
\affiliation{Laboratory for Topological Quantum Matter and Spectroscopy (B7), Department of Physics, Princeton University, Princeton, New Jersey 08544, USA}

\author{Xian P. Yang}
\affiliation{Laboratory for Topological Quantum Matter and Spectroscopy (B7), Department of Physics, Princeton University, Princeton, New Jersey 08544, USA}

\author{Zurab Guguchia}
\affiliation{Laboratory for Topological Quantum Matter and Spectroscopy (B7), Department of Physics, Princeton University, Princeton, New Jersey 08544, USA}
\affiliation{Laboratory for Muon Spin Spectroscopy, Paul Scherrer Institute, Villigen PSI, Switzerland}

\author{Stepan S. Tsirkin}
\affiliation{Department of Physics, University of Zurich, Winterthurerstrasse 190, 8057 Zurich, Switzerland}

\author{Jia-Xin Yin}
\affiliation{Laboratory for Topological Quantum Matter and Spectroscopy (B7), Department of Physics, Princeton University, Princeton, New Jersey 08544, USA}

\author{Praveen Vir}
\affiliation{Max Planck Institute for Chemical Physics of Solids, N\"othnitzer Stra{\ss}e 40, 01187 Dresden, Germany}

\author{Gohil S. Thakur}
\affiliation{Max Planck Institute for Chemical Physics of Solids, N\"othnitzer Stra{\ss}e 40, 01187 Dresden, Germany}
\affiliation{Faculty of Chemistry and Food Chemistry, Technische Universitat, 01069 Dresden, Germany}

\author{Songtian S. Zhang}
\affiliation{Laboratory for Topological Quantum Matter and Spectroscopy (B7), Department of Physics, Princeton University, Princeton, New Jersey 08544, USA}

\author{Junyi Zhang}
\affiliation{Department of Physics, Princeton University, Princeton, New Jersey 08544, USA}

\author{Konstantine Kaznatcheev}
\affiliation{National Synchrotron Light Source II, Brookhaven National Laboratory, Upton, New York 11973, USA}

\author{Guangming Cheng}
\affiliation{Princeton Institute for Science and Technology of Materials, Princeton University, Princeton, New Jersey, 08544, USA}

\author{Guoqing Chang}
\affiliation{Division of Physics and Applied Physics, School of Physical and Mathematical Sciences, Nanyang Technological University, 21 Nanyang Link, 637371, Singapore}

\author{Daniel Multer}
\affiliation{Laboratory for Topological Quantum Matter and Spectroscopy (B7), Department of Physics, Princeton University, Princeton, New Jersey 08544, USA}

\author{Nana Shumiya}
\affiliation{Laboratory for Topological Quantum Matter and Spectroscopy (B7), Department of Physics, Princeton University, Princeton, New Jersey 08544, USA}

\author{Maksim Litskevich}
\affiliation{Laboratory for Topological Quantum Matter and Spectroscopy (B7), Department of Physics, Princeton University, Princeton, New Jersey 08544, USA}

\author{Elio Vescovo}
\affiliation{National Synchrotron Light Source II, Brookhaven National Laboratory, Upton, New York 11973, USA}

\author{Timur K. Kim}
\affiliation{Diamond Light Source, Didcot OX11 0DE, UK}

\author{Cephise Cacho}
\affiliation{Diamond Light Source, Didcot OX11 0DE, UK}

\author{Nan Yao}
\affiliation{Princeton Institute for Science and Technology of Materials, Princeton University, Princeton, New Jersey, 08544, USA}

\author{Claudia Felser}
\affiliation{Max Planck Institute for Chemical Physics of Solids, N\"othnitzer Stra{\ss}e 40, 01187 Dresden, Germany}

\author{Titus Neupert}
\affiliation{Department of Physics, University of Zurich, Winterthurerstrasse 190, 8057 Zurich, Switzerland}

\author{M. Zahid Hasan} \email{mzhasan@princeton.edu}
\affiliation{Laboratory for Topological Quantum Matter and Spectroscopy (B7), Department of Physics, Princeton University, Princeton, New Jersey 08544, USA}
\affiliation{Princeton Institute for Science and Technology of Materials, Princeton University, Princeton, New Jersey, 08544, USA}
\affiliation{Materials Sciences Division, Lawrence Berkeley National Laboratory, Berkeley, CA 94720, USA}


\begin{abstract}
The manipulation of topological states in quantum matter is an essential pursuit of fundamental physics and next-generation quantum technology. Here we report the magnetic manipulation of Weyl fermions in the kagome spin-orbit semimetal \s, observed by high-resolution photoemission spectroscopy. We demonstrate the exchange collapse of spin-orbit-gapped ferromagnetic Weyl loops into paramagnetic Dirac loops under suppression of the magnetic order. We further observe that topological Fermi arcs disappear in the paramagnetic phase, suggesting the annihilation of exchange-split Weyl points. Our findings indicate that magnetic exchange collapse naturally drives Weyl fermion annihilation, opening new opportunities for engineering topology under correlated order parameters.
\end{abstract}

\date{\today}
\maketitle

Quantum magnets exhibiting electronic topology are attracting considerable interest for the magnetic manipulation of Weyl and Dirac quasiparticles, as well as their topological surface states \cite{news_daSilvaNeto,RMPWeylDirac_Armitage,ARCMP_me,ReviewNatPhys_Fiete,HighFold_NatRevMat, Review_KeimerMoore,ReviewQuantumMaterials_Hsieh}. To date, spectroscopic signatures of electronic topological ground states have been observed in several magnetic semimetals, comprising magnetic Weyl loops \cite{Co2MnGa_me}; Weyl points \cite{Co3Sn2S2_HechangLei,Co3Sn2S2_Enke,Co3Sn2S2_Beidenkopf,Co3Sn2S2_YulinChen,PrAlGe_DanDan}; and massive Dirac fermions \cite{Fe3Sn2_Checkelsky}. In parallel, the magnetic manipulation of Weyl and Dirac fermions has been extensively explored in transport \cite{Transport_ARCMP,Na3Bi_XiongJun,TaP_Chenglong,Co2MnGa_Kaustuv,Co2MnGa_Nakatsuji,Mn3Sn_Nakatsuji}. However, direct spectroscopic observation of magnetic control of topology remains challenging. Demonstrating coherent evolution of topological quasiparticles under varying magnetic order, such as the annihilation of Weyl points, offers the possibility to directly verify fundamental notions of topological band theory \cite{RMPWeylDirac_Armitage,ARCMP_me,TINI_BurkovBalents,PyrochloreWeyl_Vishwanath}. Furthermore, novel transport and optical effects are enabled by tuning the relative energies of Weyl loops and points \cite{WeylBrokenPT_Burkov,ImbalancedLineNode_Zuyzin,Gyrotropic_Souza}, controlling their positions relative to the Fermi level \cite{OpticalWeyl_Qiong,Burch_PhotovoltaicWeyl,Heinonen_SpinChargeWeyl,Fe3Ga_Nakatsuji,Co2MnGa_me} and switching on/off their topological surface states \cite{PlasmonWeyl_Song, PlasmonWeyl_Jafari}.

We have investigated magnetic modulation of topological semimetallic states in a range of materials by spectroscopy, including Fe$_3$Sn$_2$, Co$_2$MnGa, PrAlGe, Fe$_3$GeTe$_2$, TbMn$_6$Sn$_6$ and Co$_3$Sn$_2$S$_2$ \cite{Fe3Sn2_Jiaxin,Co2MnGa_me,PrAlGe_DanDan,TbMn6Sn6_Jiaxin,Co3Sn2S2_Jiaxin}. Some of these materials exhibit high magnetic transition temperatures $> 600$ K, so that thermal broadening may fundamentally overwhelm magnetic evolution of the Weyl or Dirac state \cite{Co2MnGa_me,Fe3Sn2_Checkelsky}. Other systems, such as PrAlGe, exhibit low transition temperatures of $\sim 10$ K, associated with only small magnetic perturbations to the electronic structure \cite{PrAlGe_DanDan}. Even in materials such as Fe$_3$GeTe$_2$, with intermediate $T_{\rm C} = 230$ K, the thermal evolution appears to be dominated by a suppression of quasiparticle lifetime, without significant coherent evolution of the dispersion \cite{Fe3GeTe2_Kim_Pohang,Fe3GeTe2_XinchunLai,Fe3GeTe2_YulinChen}. Using newly available high-quality single crystals combined with state-of-the-art variable-temperature photoemission spectroscopy, we have found that a large and previously overlooked energy shift of a topological spin-orbit gapped Weyl loop occurs in Co$_3$Sn$_2$S$_2$ across $T_{\rm C} = 176$ K \cite{Co3Sn2S2_QiunanXu,Co3Sn2S2_Zurab,Co3Sn2S2_Takahashi,Co3Sn2S2_XianggangQiu,Co3Sn2S2_Rossi,Co3Sn2S2_temp_DFLiu}. This shift takes place together with a magnetic exchange gap collapse that suggests a ferromagnetic Weyl to paramagnetic Dirac loop transition on raising temperature. This transition is further accompanied by the removal of candidate topological Fermi arc surface states and the annihilation of Weyl points.

Materials with inversion symmetry, mirror symmetry and ferromagnetism provide a unique platform for a magnetic-topological phase transition. The ferromagnetism produces singly-degenerate spin-split bands. In the limit of weak spin-orbit coupling (SOC), mirror symmetry can then give rise to Weyl loops on mirror planes of the bulk Brillouin zone \cite{Co2MnGa_me,WeylLoopSuperconductor_Nandkishore,WeylLines_Kane,WeylDiracLoop_Nandkishore}. A Weyl loop is a closed curve along which the bands are everywhere two-fold degenerate; it is characterized by a $\pi$ Berry phase topological invariant and a linear energy-momentum dispersion everywhere along the loop. If the magnetic order is removed and no spin-splitting remains, opposite-spin partner Weyl loops naturally collapse into a Dirac loop, a closed curve along which the bands are everywhere four-fold degenerate \cite{WeylDiracLoop_Nandkishore,DiracLineNodes_WeiderKane,CaAgX_Okamoto}. Weyl loops under SOC typically gap out, concentrating a loop of Berry curvature in momentum space, leading to a giant anomalous Hall response \cite{Co3Sn2S2_Enke,Co2MnGa_me}, large anomalous Nernst effect \cite{Co3Sn2S2_ZeroFieldNernst,Fe3Ga_Nakatsuji}, large optical Hall conductivity \cite{Co3Sn2S2_Takahashi} and other exotic response. Under SOC, Weyl loops may also leave behind some discrete number of Weyl points. By contrast, Dirac points are generically unstable under inversion and time-reversal symmetry \cite{Gapless_Murakami}, so that Dirac loops under SOC gap out fully. As a result, in this scenario upon magnetic exchange collapse the Weyl points generically annihilate.

\begin{figure}
\centering
\includegraphics[width=16cm,trim={0in 0in 0in 0in},clip]{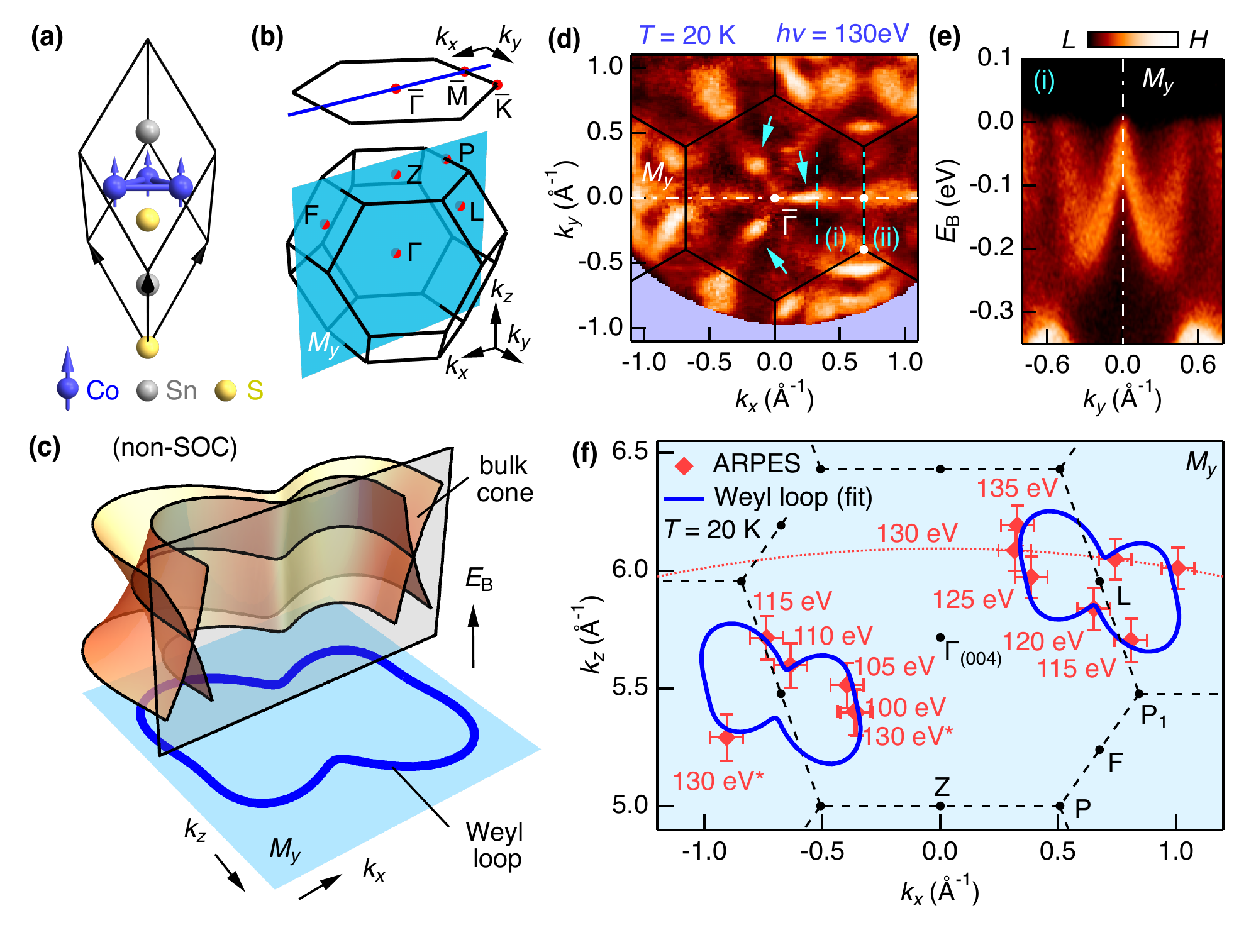}
\caption{\label{intro} {\bf Topological magnetic Weyl loop.} \cpana\ Primitive unit cell of ferromagnetic \s, with mirror symmetry. \cpanb\ Bulk and $(001)$ surface Brillouin zones with bulk mirror plane ($M_y$, cyan) and several high-symmetry points (red). \cpanc\ In the absence of spin-orbit coupling (SOC), the combination of mirror symmetry and ferromagnetism generically gives rise to Weyl loops, which live in the mirror planes of the bulk Brillouin zone. A Weyl loop exhibits a ring of band crossings along a closed curve in momentum space (blue loop) with a linear cone dispersion on any energy-momentum slice through the loop. Under SOC, the Weyl loop typically gaps, possibly leaving behind Weyl points. \cpand\ ARPES Fermi surface at $T = 20$ K and photon energy $h\nu = 130$ eV, exhibiting multiple dot features (cyan arrows) on the mirror planes ($\bar{\Gamma}-\bar{M}$). \cpane\ Cone dispersion at the Fermi level on an energy-momentum spectrum through the dot feature (Cut i). \cpanf\ Collecting cone dispersions for a range of $h\nu$ suggests a loop of band crossings (red diamonds) living in $M_y$ and encircling the bulk L point (Fig. S4, S5). Different $h\nu$ sample different out-of-plane $k_z$ momenta; representative example shown for $130$ eV (dotted red curve). The crossing points can be fit by a low-order polar coordinate Fourier decomposition around the L point (blue curve \cite{SM}), mapping out the trajectory of the Weyl loop.}
\end{figure}

\s\ crystallizes in space group $R\bar{3}2/m$ (No. 166), with dihedral point group $D_{3d}$, which includes inversion symmetry and three mirror planes (Fig. \ref{intro}\pana, S3). The system is ferromagnetic, with Curie temperature $T_{\rm C} = 176$ K \cite{Co3Sn2S2_Sobany,Co3Sn2S2_Weihrich}. Keeping in mind the mirror symmetry and ferromagnetic order, we explore our \s\ samples by ARPES at 20 K. Measuring with incident photon energy $h\nu = 130$ eV, we observe point-like electronic structures on $M_y$ (cyan arrows, Fig. \ref{intro}\pand; the mirror planes correspond to $\bar{\Gamma}-\bar{M}$). On cuts along $k_y$ through the point-like features, we observe cone dispersions straddling the mirror plane $M_y$ (Figs. \ref{intro}\pane). On an energy-momentum cut along $k_x$, within the mirror plane, we again observe cone-like dispersions (Fig. S4). The observation of cone dispersions along both $k_x$ and $k_y$, coming together at point Fermi surfaces, suggests a set of band crossings living in the momentum-space mirror plane. To systematically understand the evolution of the band crossings along the out-of-plane momentum-space direction $k_z$ we acquire analogous datasets at a range of photon energies, from $h\nu = 100$ to $135$ eV (Fig. S5). We find that the cones persist in $h\nu$, with crossing points consistently on the $M_y$ plane, but at varying $(k_x, k_z)$ coordinates (red diamonds, Fig. \ref{intro}\panf). Taken together, these crossing points appear to form an extended nodal electronic state encircling the L point of the bulk Brillouin zone, suggesting the observation of a bulk loop node in \s. Since the system is ferromagnetic with generically singly-degenerate bands, we interpret this loop node as a Weyl loop (Fig. \ref{intro}\panc). To extract the complete trajectory of the loop, we fit the ARPES locations of the cone dispersions to a low-order polar Fourier decomposition around the L point of the bulk Brillouin zone (blue loop, Fig. \ref{intro}\panf; see Supplemental Material \cite{SM} for fitting parameter values). In this way we extract the full momentum-space trajectory of the Weyl loop from photoemission data alone.

Next we explore the evolution of the Weyl loop with temperature, focusing on Cut (i). We systematically cycle the temperature of our samples from 20 K to 290 K and back to 20 K, moving across $T_{\rm C} = 176$ K. On raising the temperature, we observe a dramatic evolution of the Weyl cone on a large energy scale of $\sim 0.1$ eV (Figs. \ref{tdep}\pana, \panb; S8), with the cone appearing to recede above \f. We next assemble the momentum distribution curves (MDCs) of Cut (i) at \f\ for all temperatures (Fig. \ref{tdep}\panc). Upon cycling the temperature, we observe a prominent and reversible evolution of the Weyl cone across \tc, consistent with a magnetic phase transition. For further insight, we examine additional spectra on Cut (ii), obtained during the course of the same measurement, and we consider a set of deep bands $\sim 0.3$ eV below \f, which are predominantly formed from the same exchange-split Co $3d$ $a_{1g}$ and $e_g$ manifolds as the Weyl loop (Figs. \ref{tdep}\pand; S11). At 20 K, these deep valence bands exhibit clear splitting, consistent with the material's ferromagnetic order. Upon raising the temperature, the splitting appears to vanish and these deep bands collapse together, suggesting a paramagnetic state with spin-degenerate bands. By examining the evolution of the deep bands, we circumvent the limitations of the photoemission \f\ cut-off and observe direct signatures of a prominent magnetic exchange gap collapse across \tc\ in \s.

To relate the Weyl loop temperature evolution to the magnetic exchange gap collapse, we consider more carefully the interplay between topology and ferromagnetism. In \ab\ calculation, in the absence of SOC and in the ferromagnetic state, the Weyl loop arises as a crossing of two spin-majority bands, with a spin-minority partner Weyl loop above the Fermi level (schematic blue and green loops, Figs. \ref{tdep}\pane; S6). In a non-magnetic \ab\ calculation, the exchange gap vanishes and these two Weyl loops coincide, forming a spinless loop crossing---a Dirac loop (purple loop). Comparing the \ab\ calculations with ARPES, we find that the magnetic Weyl and non-magnetic Dirac nodes exhibit overall agreement with the ferromagnetic and paramagnetic spectra, respectively (magenta traces, Fig. \ref{tdep}\panb). Note that including SOC in our \ab\ results does not alter this interpretation, although the expected gap appears in both loop nodes (blue traces). The observation that the loop recedes above \f\ on increasing temperature is also consistent with maintenance of charge balance in the spin-degenerate electronic structure, further indicating a paramagnetic Dirac loop. We further reduced the magnetic moment in our samples via nickel (Ni) doping, and again observed a persistent loop node electronic structure despite suppression of the ferromagnetism (Fig. S1). Taken together, our systematic ARPES spectra and \ab\ calculations suggest that we have observed the collapse of two opposite-spin ferromagnetic Weyl loops into a paramagnetic Dirac loop.

To quantitatively characterize the Weyl loop collapse with temperature, we perform a Lorentzian fit of energy distribution curves (EDCs) through the extremum of the Weyl loop band (Cut(i), dotted line, Fig. \ref{tdep}\pana). The extracted Weyl band extremum exhibits a clear evolution upward in \eb\ as the temperature increases, 20 K $\rightarrow$ 250 K, consistent with exchange gap collapse (Fig. \ref{tdep}\pand). We further extract the exchange gap $\Delta(T)$ on EDCs through the deep bands (Cut (ii), dotted line, Fig. \ref{tdep}\panc) and compared the resulting $\Delta(T)$ with the magnetization $M(T)$ as measured by a SQUID. For $T < T_{\rm C}$ we find that the exchange splitting tracks $M(T)$. For $T > T_{\rm C}$ we no longer observe an exchange splitting within our spectral linewidth, consistent with the absence of magnetization. Remarkably, the observed exchange gap and Weyl band shift are both $\sim 0.12$ eV, suggesting a complete collapse of the opposite-spin partner Weyl loops across \tc\ and the formation of a spin-degenerate paramagnetic Dirac loop.

To further explore the paramagnetic Dirac loop we park our apparatus at 220 K, well into the paramagnetic phase. At a range of $h\nu$ we observe characteristic point-like iso-energy contours on $M_y$ and related mirror planes (Figs. \ref{dirac}\pana, \panc; S7). Energy-momentum spectra through these point-like contours further exhibit cone-like spectral weight straddling $M_y$, indicative of Dirac loop cone dispersions above \f\ (Figs. \ref{dirac}\panb, \pand). The presence of multiple cone features straddling $M_y$ at a range of $h\nu$ again suggests an extended nodal electronic structure confined to the mirror plane. Since we are in the paramagnetic phase with generically spin-degenerate bands, we interpret these candidate band crossings as four-fold degenerate, forming a Dirac loop. By analogy with our analysis in the ferromagnetic phase, we again systematically collect the locations of all cone features observed in the paramagnetic phase for $h\nu$ from 100 to 135 eV and experimentally extract the full momentum-space trajectory of the Dirac loop (red diamonds, Fig. \ref{dirac}\pane). \Ab\ calculations of the Weyl and Dirac loops in the ferromagnetic and non-magnetic states also agree with the experimentally-observed trajectories (Fig. S2). A loop node electronic structure persisting into the paramagnetic phase of \s\ again suggests the observation of a ferromagnetic Weyl to paramagnetic Dirac loop collapse.

\begin{figure}
\centering
\includegraphics[width=17cm,trim={0in 0in 0in 0in},clip]{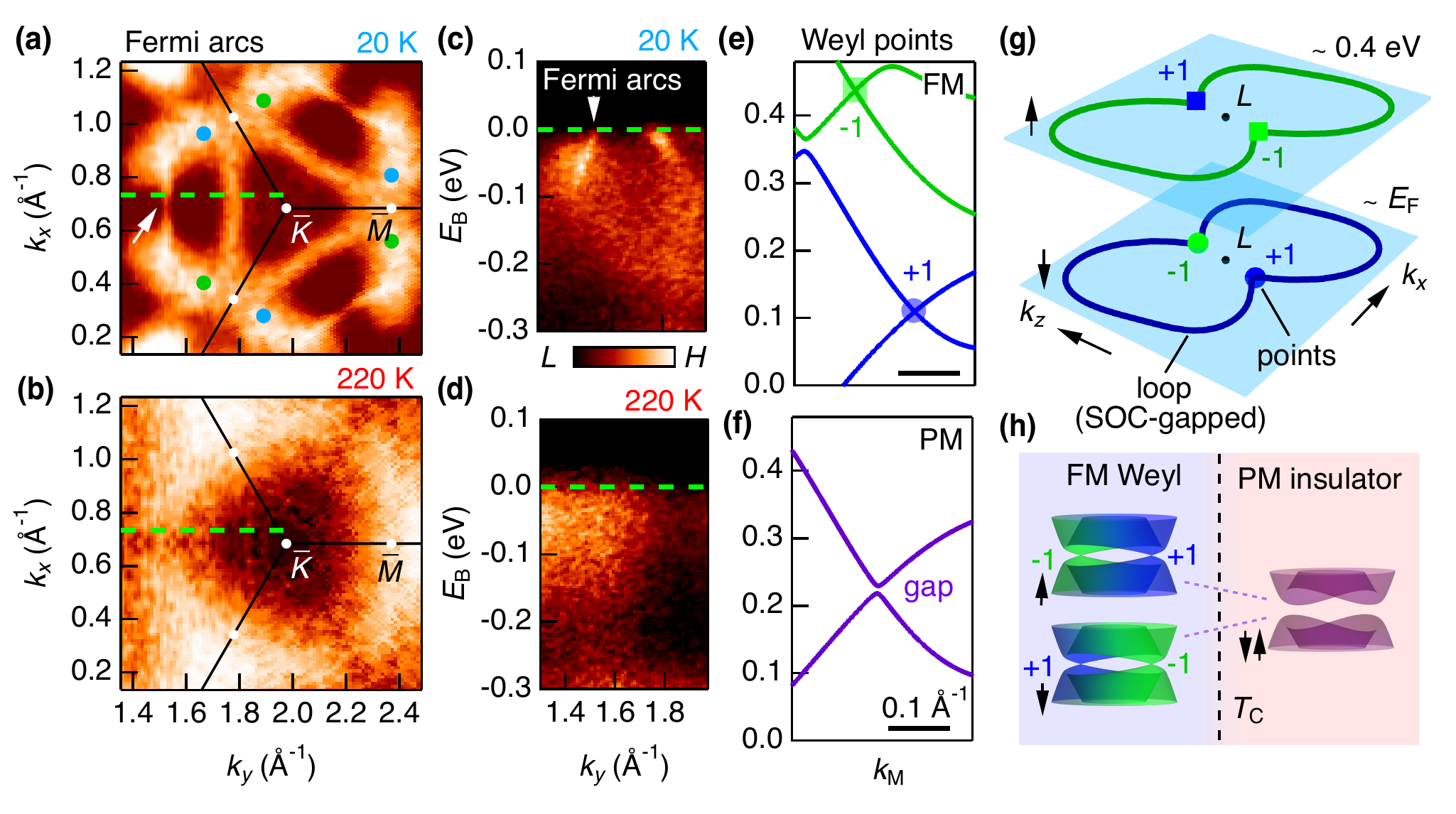}
\caption{\label{annihilation} {\bf Evidence for Fermi arc and Weyl point annihilation.} \cpana\ Fermi surface acquired at $T = 20$ K, $h\nu = 130$ eV, exhibiting candidate Fermi arc surface states (white arrow) near the Weyl point locations, as predicted by \ab\ calculations (blue, green circle; Fig. S9). Data symmetrized for clarity. \cpanb\ Analogous Fermi surface at 220 K, with no signature of Fermi arcs. \cpanc\ Energy-momentum cut through the candidate Fermi arcs at 20 K (green dotted line, (\pana)). \cpand\ Analogous cut at 220 K. \cpane\ Calculation in the ferromagnetic state, in the presence of SOC, slicing through a pair of exchange-split Weyl points of opposite chirality. \cpanf\ Analogous calculation, non-magnetic state. The two partner Weyl points annihilate, opening a gap of $12$ meV. \cpang\ Schematic of the ferromagnetic Weyl loop (non-SOC) and Weyl point (SOC) configuration. \cpanh\ Schematic phase diagram: spin-orbit gapped ferromagnetic Weyl loops collapse to a paramagnetic Dirac loop across \tc. Concurrently, the exchange-split Weyl points annihilate.}
\end{figure}

We next consider the fine structure of the Weyl loop collapse associated with spin-orbit coupling (SOC). In \s, \ab\ calculations along with ARPES and STM investigations suggest that each Weyl loop under SOC produces two Weyl points above the Fermi level, with signatures of topological Fermi arc surface states extending below the Fermi level. \cite{Co3Sn2S2_Enke,Co3Sn2S2_Beidenkopf,Co3Sn2S2_QiunanXu,Co3Sn2S2_HechangLei,Co3Sn2S2_Zurab,Co3Sn2S2_YulinChen,Co3Sn2S2_Takahashi,Co3Sn2S2_XianggangQiu}. Our observation of a Weyl to Dirac loop transition naturally motivates investigation of Fermi arc and Weyl point annihilation across $T_{\rm C}$. At 20 K, we observe sharp arc-shaped states near the expected Weyl points, consistent with topological Fermi arcs in \ab\ calculation (Figs. \ref{annihilation}\pana, \panc; S9). At 220 K these Fermi arcs vanish, leaving behind bulk pockets broadly consistent with the low temperature spectra (Figs. \ref{annihilation}\panb, \pand). The disappearance of the Fermi arcs above \tc\ provides evidence for the annihilation of Weyl points in the paramagnetic phase. To further characterize this annihilation, we consider the \ab\ band structure under ferromagnetic order on a momentum-space path connecting a pair of exchange-split Weyl points (Fig. \ref{annihilation}\pane). Upon exchange collapse, these partner Weyl points come together and annihilate, opening a gap (Fig. \ref{annihilation}\panf).

Our systematic variable-temperature ARPES experiments suggest that pairs of ferromagnetic Weyl loops collapse into paramagnetic Dirac loops across \tc\ in \s\ (Figs. \ref{annihilation}\pang, \panh). Taken together with \ab\ calculations, our results additionally provide evidence for the annihilation of Fermi arcs and Weyl points concomitant with this transition. Our findings suggest a general mechanism for Weyl fermion annihilation, where the annihilation is driven by magnetic exchange gap collapse and takes place predominantly along the energy axis, rather than in momentum \cite{TaAs_DFT_ShinMing,ARCMP_me,TINI_BurkovBalents}. This novel mechanism should occur naturally in many quantum magnets and motivates exploration of the rich topological evolution associated with the onset of magnetic order. Such interplay between magnetism and topology may also pave the way to magnetic design of correlated topological states with exotic transport and optical response.

\section{Acknowledgments}

The authors thank D. Lu and M. Hashimoto at Beamline 5-2 of the Stanford Synchrotron Radiation Lightsource (SSRL) at the SLAC National Accelerator Laboratory, CA, USA for support. The authors thank Diamond Light Source for access to Beamline I05 (SI17924, SI19313). This research used Beamline 21-ID-1 (ESM-ARPES) of the National Synchrotron Light Source II, a U.S. Department of Energy (DOE) Office of Science User Facility operated for the DOE Office of Science by Brookhaven National Laboratory under Contract No. DE-SC0012704. The authors also acknowledge use of Princeton University’s Imaging and Analysis Center, which is partially supported by the Princeton Center for Complex Materials (PCCM), a National Science Foundation (NSF)-MRSEC program (DMR-2011750). Use of the Stanford Synchrotron Radiation Lightsource (SSRL), SLAC National Accelerator Laboratory, is supported by the U.S. Department of Energy, Office of Science, Office of Basic Energy Sciences, under Contract No. DE-AC02-76SF00515. T. A. C. acknowledges the support of the National Science Foundation Graduate Research Fellowship Program (DGE-1656466). T. N. and S. S. T. acknowledge support from the European Union Horizon 2020 Research and Innovation Program (ERC-StG-Neupert-757867-PARATOP). S. S. T. also acknowledges support from the Swiss National Science Foundation (Grant No. PP00P2-176877). X. L. acknowledges financial support from the China Scholarship Council. G. C. would like to acknowledge the support of the National Research Foundation, Singapore under its NRF Fellowship Award (NRF-NRFF13-2021-0010) and the Nanyang Assistant Professorship grant from Nanyang Technological University. M. Z. H. acknowledges visiting scientist support at Berkeley Lab (LBNL) during the early phases of this work. Work at Princeton University was supported by the Gordon and Betty Moore Foundation
(Grants No. GBMF4547 and No. GBMF9461; M. Z. H.). The ARPES and theoretical work were supported by the
United States Department of Energy (US DOE) under the Basic Energy Sciences programme (Grant No. DOE/BES
DE-FG-02-05ER46200; M. Z. H.). G. S. T. thanks the Würzburg-Dresden Cluster of Excellence on Complexity
and Topology in Quantum Matter – ct.qmat (EXC 2147) for postdoctoral funding. C. F. acknowledges the DFG through
SFB 1143 (project ID. 247310070) and the WürzburgDresden Cluster of Excellence on Complexity and Topology in Quantum Matter ct.qmat (EXC2147, project ID. 39085490).


\clearpage

\clearpage
\begin{figure}
\centering
\includegraphics[width=17cm,trim={0in 0in 0in 0in},clip]{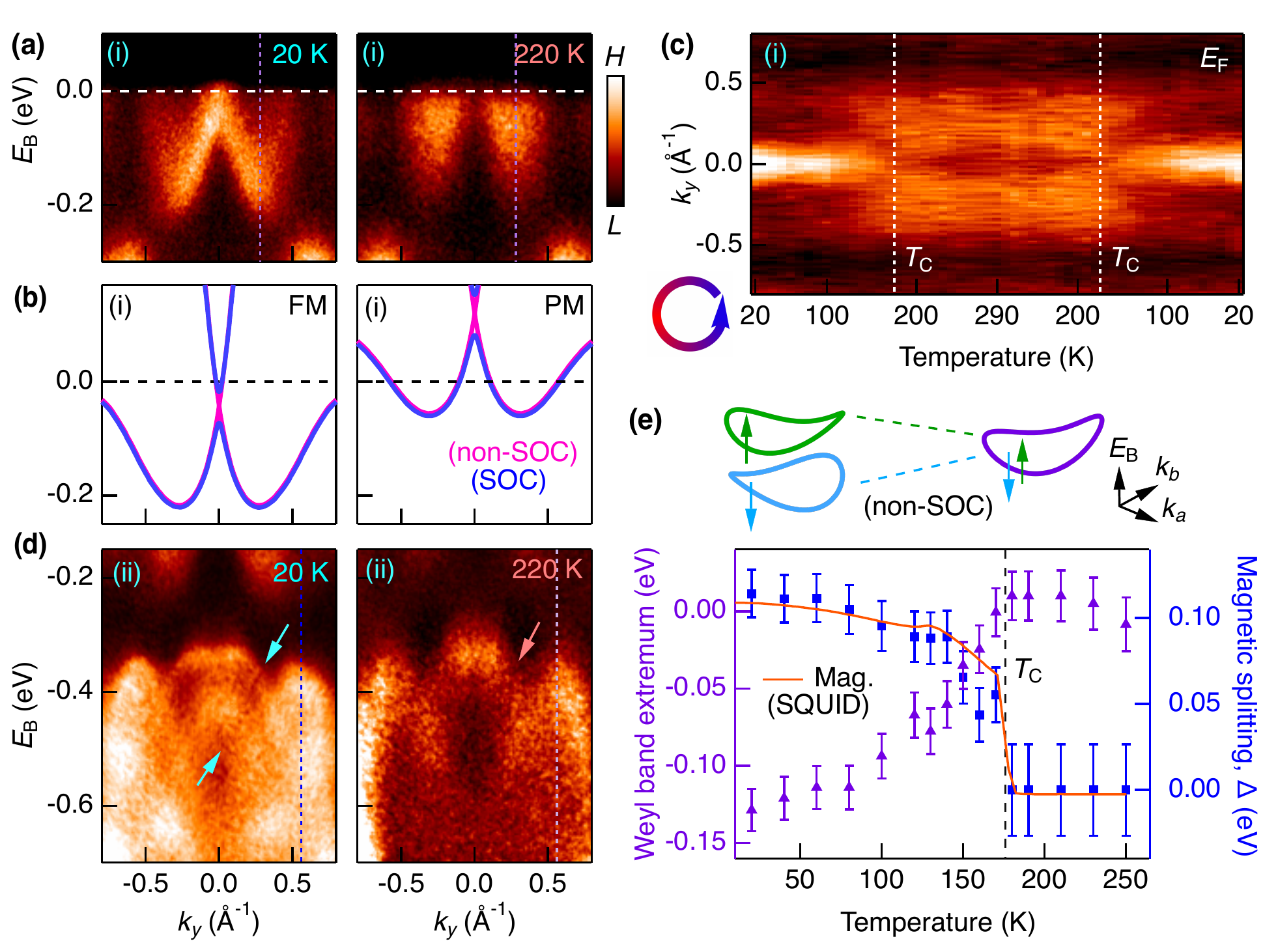}
\end{figure}
\begin{figure}
\caption{\label{tdep} {\bf Magnetic Weyl to Dirac loop collapse.} \cpana\ Cut (i) at 20 K and 220 K, with \cpanb\ corresponding \ab\ calculation. Left: calculation in the ferromagnetic state through the Weyl loop, without SOC (magenta) and with SOC (blue). Right: calculation in the non-magnetic state through the Dirac loop. \cpanc\ Momentum distribution curves (MDCs) of Cut (i) at the Fermi level for the full temperature cycle, 20 K $\rightarrow$ 290 K $\rightarrow$ 20 K. \cpand\ Cut (ii), defined in Fig. \ref{intro}\pand, exhibiting clear splittings in deeper energy bands at 20 K (left), which collapse at 220 K (right). \cpane\ Energy extremum of the Weyl loop band, extracted from the temperature dependence on Cut (i), obtained by Lorentzian fitting of energy distribution curves (EDCs, dotted lines in (\pana)). Also, the magnetic exchange splitting as a function of temperature, obtained from Cut (ii) by Lorentzian fitting to EDCs (dotted lines in (\pand)) and compared with the magnetization $M(T)$. Cartoon: exchange gap collapse of two opposite-spin Weyl loop partners (blue and green) into a single Dirac loop (purple).}
\end{figure}

\clearpage
\begin{figure}
\centering
\includegraphics[width=10cm,trim={0in 0in 0in 0in},clip]{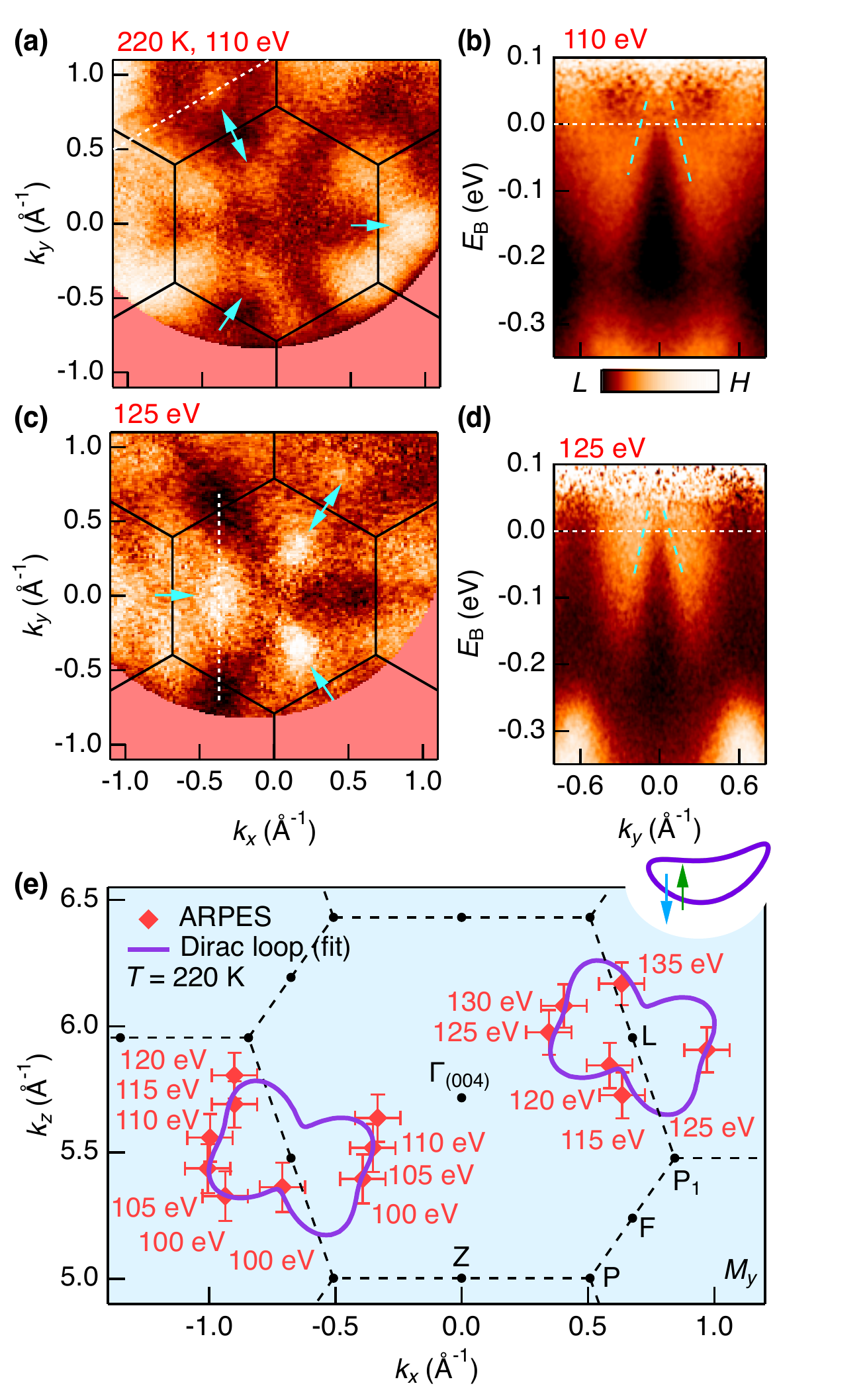}
\end{figure}
\begin{figure}
\caption{\label{dirac} {\bf Paramagnetic Dirac loop.} \cpana\ ARPES iso-energy contour slightly above \f, at $h\nu = 110$ eV, acquired at 220 K, exhibiting point-like features (cyan arrows) along $\bar{\Gamma}-\bar{M}$ (corresponding to $M_y$ and the symmetry-related mirror planes). \cpanb\ Energy-momentum cut through the point-like feature, exhibiting cone-like spectral weight (cut location: white line in (\pana)). \cpanc, \cpand\ Analogous to (\pana), (\panb), at 125 eV. \cpane\ Locations of cones observed for all $h\nu$ (red diamonds, Fig. S7). Cones on symmetry-related mirror planes are plotted all together in a single momentum-space mirror plane $M_y$. Data points fit to a low-order polar coordinate Fourier decomposition around L (purple curves), mapping out the trajectory of the Dirac loop.}
\end{figure}

\clearpage

\end{document}


\title{Supplemental Material for:\\ Signatures of Weyl fermion annihilation\\ in a correlated kagome magnet}

\author{Ilya Belopolski$^*$}

\author{Tyler A. Cochran$^*$}

\author{Xiaoxiong Liu$^*$}

\author{Zi-Jia Cheng$^*$}

\author{Xian P. Yang}

\author{Zurab Guguchia}

\author{Stepan S. Tsirkin}

\author{Jia-Xin Yin}

\author{Praveen Vir}

\author{Gohil S. Thakur}

\author{Songtian S. Zhang}

\author{Junyi Zhang}

\author{Konstantine Kaznatcheev}

\author{Guangming Cheng}

\author{Guoqing Chang}

\author{Daniel Multer}

\author{Nana Shumiya}

\author{Maksim Litskevich}

\author{Elio Vescovo}

\author{Timur K. Kim}

\author{Cephise Cacho}

\author{Nan Yao}

\author{Claudia Felser}

\author{Titus Neupert}

\author{M. Zahid Hasan} 

\pacs{}

\maketitle

\section{Weyl loop in magnetically-suppressed N\lowercase{i}-doped C\lowercase{o}$_3$S\lowercase{n}$_2$S$_2$}

We can more deeply explore the ferromagnetic Weyl to paramagnetic Dirac loop transition in nickel (Ni) doped \s. Our \ns\ samples show a well-behaved suppression of ferromagnetism upon increased dopant concentration, with approximately linear decay to zero of the \tc, the magnetization $M$ and the anomalous Hall conductivity $\sigma^{\rm A}_{xy}$ as the Ni level varies from $x = 0$ to 0.6 \cite{Co3Sn2S2_NCSS_Thakur}. At the same time, Ni nominally electron dopes the system. This allows photoemission measurements in a reduced-magnetization state with favorable positioning of the Fermi level, while maintaining base temperature for the measurement and consequently providing higher energy resolution. At $x = 0.35$, for $h\nu = 110$ eV, we observe clear point-like Fermi surfaces with sharp cone dispersion (Figs. \ref{Ni_SI}\pana,\panb). This cone dispersion persists for a range of $h\nu$, suggesting that the loop node electronic structure survives in the reduced-magnetization state, again indicating a magnetic Weyl to Dirac loop transition as ferromagnetism is suppressed (Fig. \ref{Ni_SI}\panc-\panf). The momentum-space trajectory of the Ni-doped loop node (green diamonds, Fig. \ref{Traj_SI}) quantitatively agrees with the Weyl and Dirac loop trajectories extracted for the undoped sample. These systematic results further support a ferromagnetic Weyl to paramagnetic Dirac loop transition.

\section{Methods}

\textit{Single crystal growth}: Single crystals of \s\ were grown by a self-flux method with the congruent composition. The stoichiometrically-weighted starting materials were put in a graphite crucible sealed in a quartz tube. The samples were heated to 1000$^{\circ}$C over 48 hours, left there for 24 hours, and then slowly cooled to 600$^{\circ}$C over one week. An annealing process was implemented at 600$^{\circ}$C for another 24 hours to produce homogeneous and well-ordered crystals. The compositions and phase structure of the samples were initially checked by energy-dispersive X-ray spectroscopy and powder X-ray diffraction, respectively.\\

\textit{Scanning transmission electron microscopy}: Thin lamellae for microstructure characterization were prepared from bulk single crystals by focused ion beam cutting using a FEI Helios NanoLab 600 dual beam system (FIB/SEM). Atomic resolution high-angle annular dark-field (HAADF) scanning transmission electron microscopy (STEM) imaging and atomic-level energy-dispersive X-ray spectroscopy (EDS) mapping were performed on a double Cs-corrected FEI Titan Cubed Themis 300 scanning/transmission electron microscope (S/TEM) equipped with an X-FEG source operated at 300 kV with a Super-X EDS system.\\

\textit{Magnetometry}: The magnetization measurement was carried out with a superconducting quantum interference device (SQUID) magnetometer (QuantumDesign) from 2 to 300 K in a magnetic field of 0.01 T applied parallel to the crystallographic $c$-axis.\\

\textit{Angle-resolved photoemission spectroscopy}: Photon-energy-dependent ARPES measurements were carried out at Beamline I05 of Diamond Light Source, Harwell Science Campus, Oxfordshire, UK using a Scienta R4000 electron analyzer with angular resolution $< 0.2^{\circ}$ and total energy resolution $< 13$ meV for all photon energies, from 100 to 135 eV, with spot size 50 $\mu$m $\times$ 50 $\mu$m \cite{Diamond_Hoesch}. The sample temperature was 8 K. Temperature-dependent ARPES measurements were carried out at Beamline 5-2 of the Stanford Synchrotron Radiation Lightsource, SLAC in Menlo Park, CA, USA using a Scienta R4000 electron analyzer, with photon energy $h\nu = 130$ eV; angular resolution $ < 0.2^{\circ}$; total energy resolution from 14 meV at 22 K to 30 meV at 280 K; and beam spot size 16 $\mu$m (vertical) $\times$ 36 $\mu$m (horizontal). Measurements on \ns\ samples were carried out at Beamline 21-ID-1 (ESM-ARPES) of the National Synchrotron Light Source II, BNL in Upton, NY, USA. Samples were cleaved $\textit{in situ}$ and measured under a vacuum of $4 \times 10^{-11}$ Torr or better for all temperatures. To characterize the Weyl and Dirac loops, it is natural to parametrize its trajectory by an angle $\omega$ in polar coordinates with L taken as the origin. The loop is then described by a function $r (\omega)$ with $2\pi$ periodicity. Crystalline inversion symmetry $P$ further requires that the dispersion remain unchanged under inversion through L, constraining the trajectory to $r (\omega + \pi) = r (\omega)$. With $\pi$ periodicity, the first three terms of the Fourier decomposition are $r (\omega) = r_0 + r_1 \cos (2\omega + \phi_1) + r_2 \cos (4\omega + \phi_2)$. By fitting to the ARPES locations of the cone dispersions, we find that the trajectory of the Weyl loop is given by $r_0 = 0.27\ {\rm \AA}^{-1}$, $r_1 = 0.12\ {\rm \AA}^{-1}$, $\phi_1 = 43^{\circ}$, $r_2 = -0.05\ {\rm \AA}^{-1}$ and $\phi_2 = 27^{\circ}$ (blue loop, Fig. 1f). In this way, we extract the full momentum-space trajectory of the Weyl and Dirac loops from photoemission data alone.\\

\noindent To understand the role of different surface terminations in our ARPES measurements, we cleaved 53 \s\ crystals and imaged the cleaved surface by scanning tunneling microscopy (STM). We scanned $\sim 10\ \mu m^2$ for every cleaved sample surface. We found that 33\% of the surfaces are disordered (without atomic lattice), 50\% are Sn surfaces, 15\% are S surfaces and 2\% are Co$_3$Sn surfaces. Moreover, our high-quality photoemission spectra of the surface states were consistently well-captured by \ab\ calculations with Sn termination, but not S termination (Figs. \ref{arc_SI}, \ref{Stermination}) \cite{Co3Sn2S2_YulinChen}. Our systematic STM, ARPES and \ab\ characterization suggests that only the Sn termination is relevant for our analysis.\\

\textit{\Ab\ calculations}: Density functional theory (DFT) calculations with the projected augmented wave (PAW) method were implemented in the Vienna \ab\ simulation package (VASP) \cite{DFT2,DFT_Efficiency_KressFurthmueller} with generalized gradient approximation (GGA) \cite{DFT4}. For the irreducible Brillouin zone, $k$-meshes of size $8 \times 8 \times 8$ were used. The Fermi level of the \ab\ calculation was also optimized to match the experimental results. The surface spectral function calculated by the WannierTools package \cite{Soluyanov_WannierTools}, using a tight-binding model generated from maximally-localized Wannier functions \cite{Yates_Wannier90}. It was observed that the Sn termination captured the essential features of the ARPES.

\renewcommand{\thefigure}{S\arabic{figure}}
\setcounter{figure}{0}

\clearpage
\begin{figure}
\centering
\includegraphics[width=12cm,trim={0in 0in 0in 0in},clip]{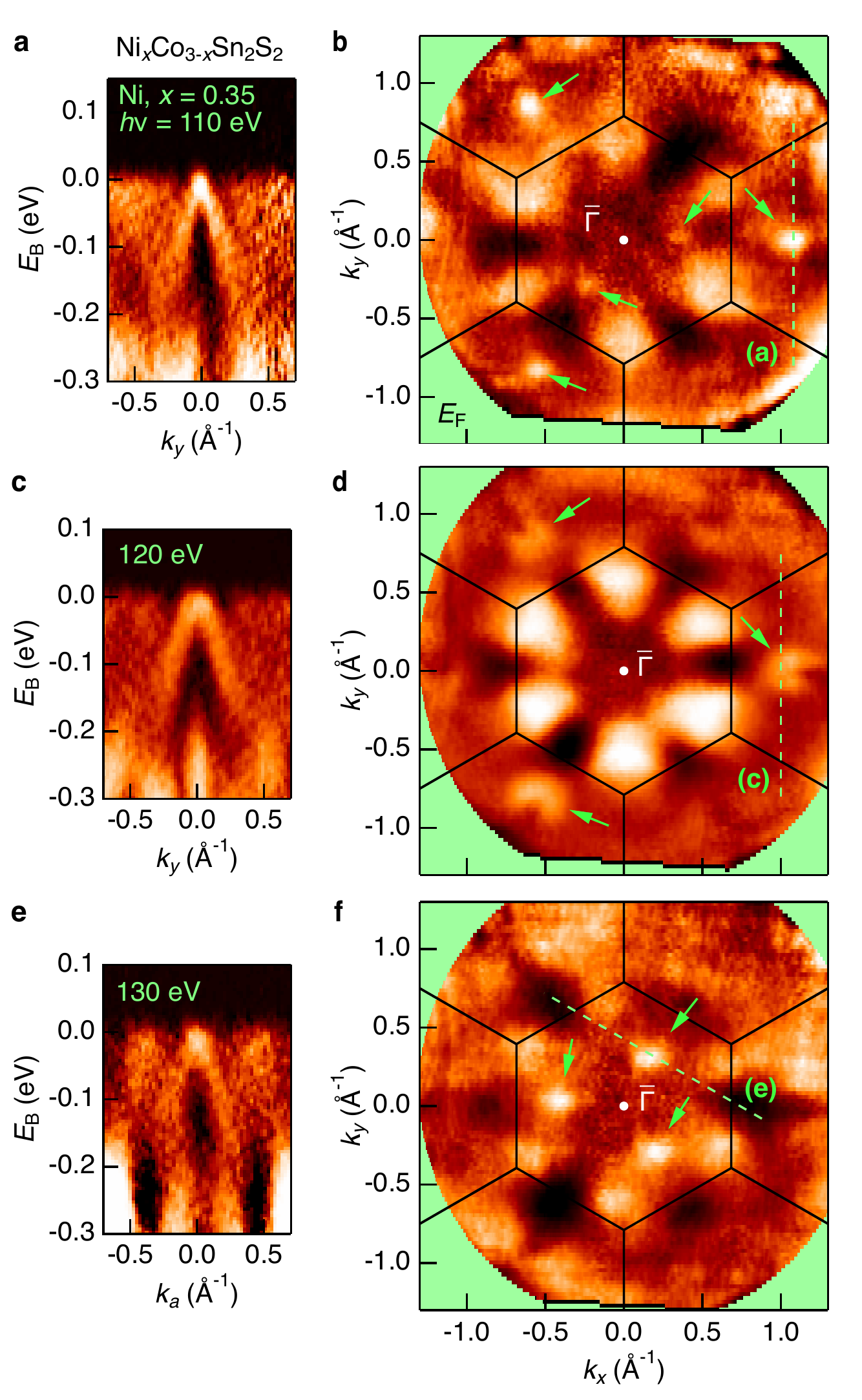}
\end{figure}
\begin{figure}
\caption{\label{Ni_SI} {\bf Weyl loop, Ni-doped \s.} \cpana\ Energy-momentum ARPES spectrum obtained on ferromagnetically-suppressed \ns\ single crystals, $x = 0.35$, at $T = 20$ K and $h\nu = 110$ eV, exhibiting a cone dispersion centered on $M_y$. \cpanb\ Corresponding Fermi surface at 110 eV, exhibiting point-like features (green arrows) on $M_y$ and symmetry-equivalent mirror planes. Spectrum in (\pand) marked by the dotted green line. \cpanc-\cpanf\ Analogous spectra acquired at $h\nu = 120$ and 130 eV. The energy-momentum spectra again show clear cone dispersions (\panc, \pane), corresponding to point-like features (green arrows, \pand, \panf) on $M_y$ and equivalent mirror planes.}
\end{figure}

\clearpage
\begin{figure}
\centering
\includegraphics[width=8cm,trim={0in 0in 0in 0in},clip]{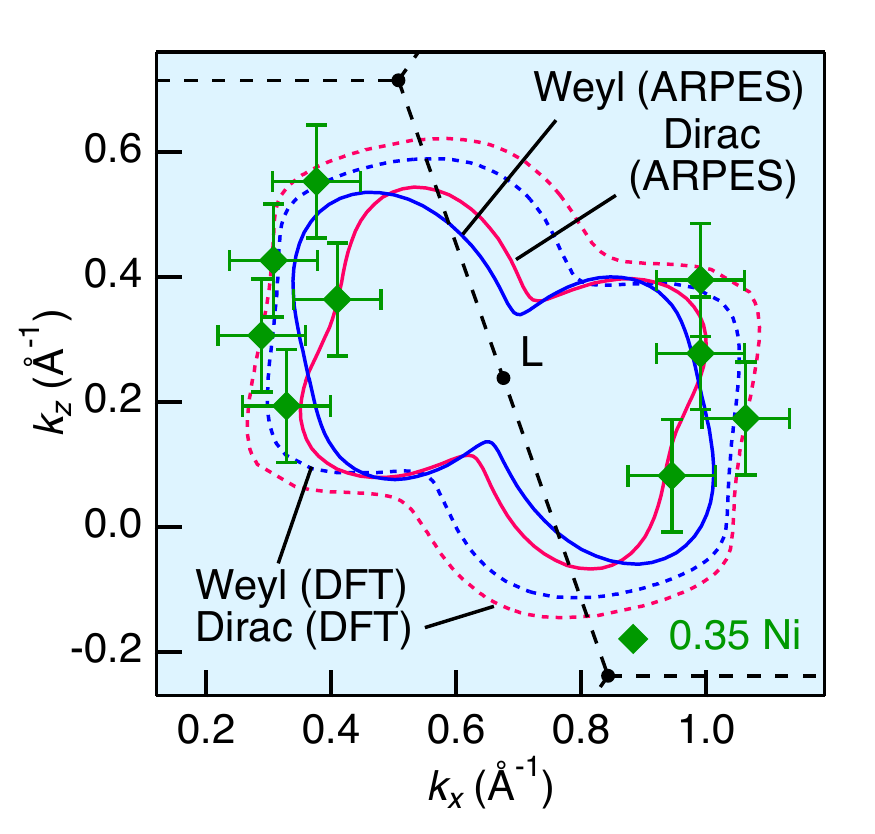}
\caption{\label{Traj_SI} {\bf Loop node trajectories.} Summary of the ferromagnetic Weyl and paramagnetic Dirac loop trajectories extracted by photoemission; the analogous \ab\ result in the ferromagnetic and non-magnetic states, in the absence of SOC; and cone dispersions observed in \ns.}
\end{figure}

\clearpage
\begin{figure}
\centering
\includegraphics[width=14cm,trim={0in 0in 0in 0in},clip]{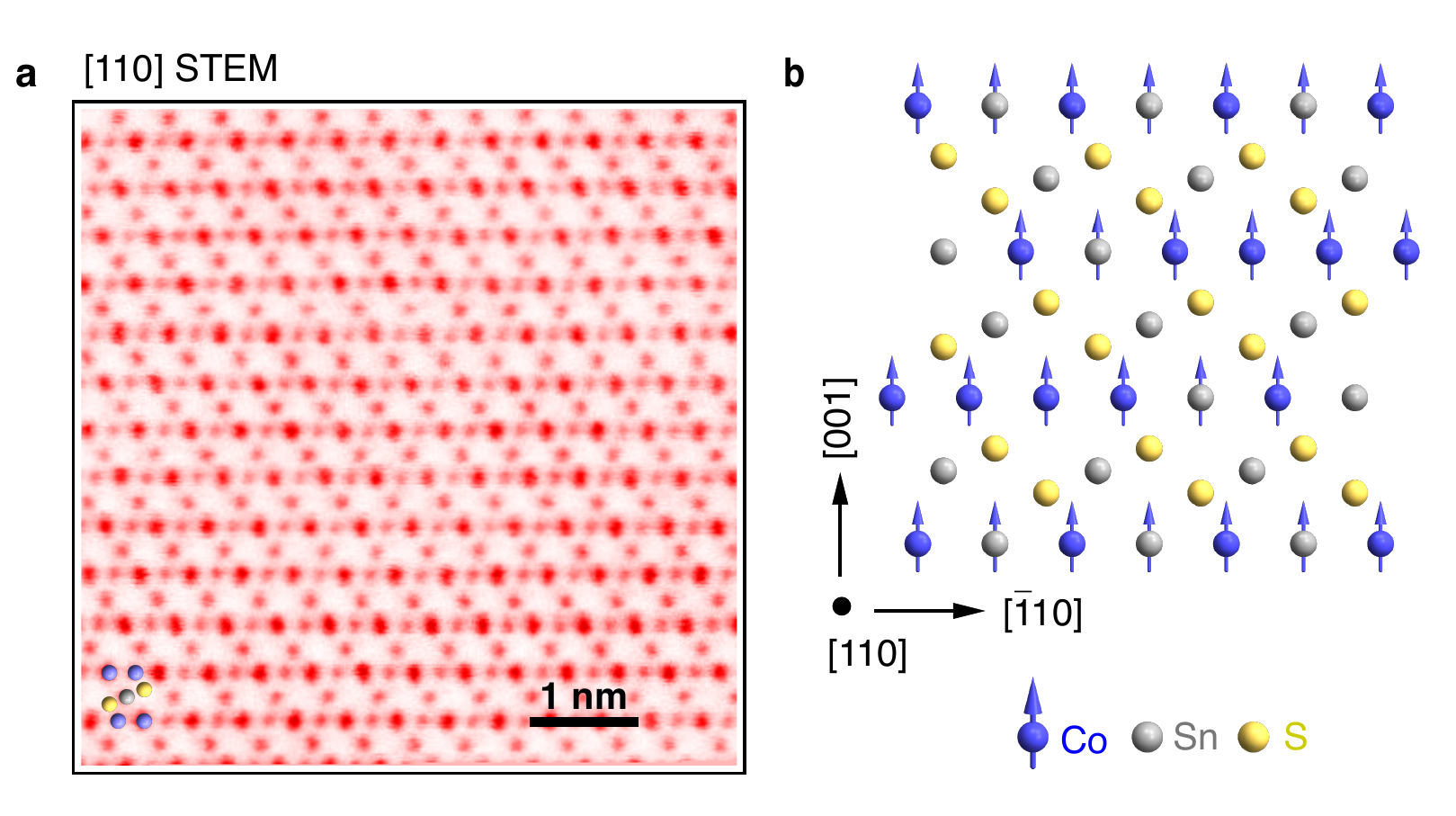}
\caption{\label{crys_SI} {\bf Crystal structure.} \cpana\ Crystal structure of \s\ along $[110]$, measured by scanning transmission electron microscopy (STEM). \cpanb\ Schematic crystal structure viewed along $[110]$, exhibiting the two-fold rotation $C_{2x}$ symmetry.}
\end{figure}

\clearpage
\begin{figure}
\centering
\includegraphics[width=10cm,trim={0in 0in 0in 0in},clip]{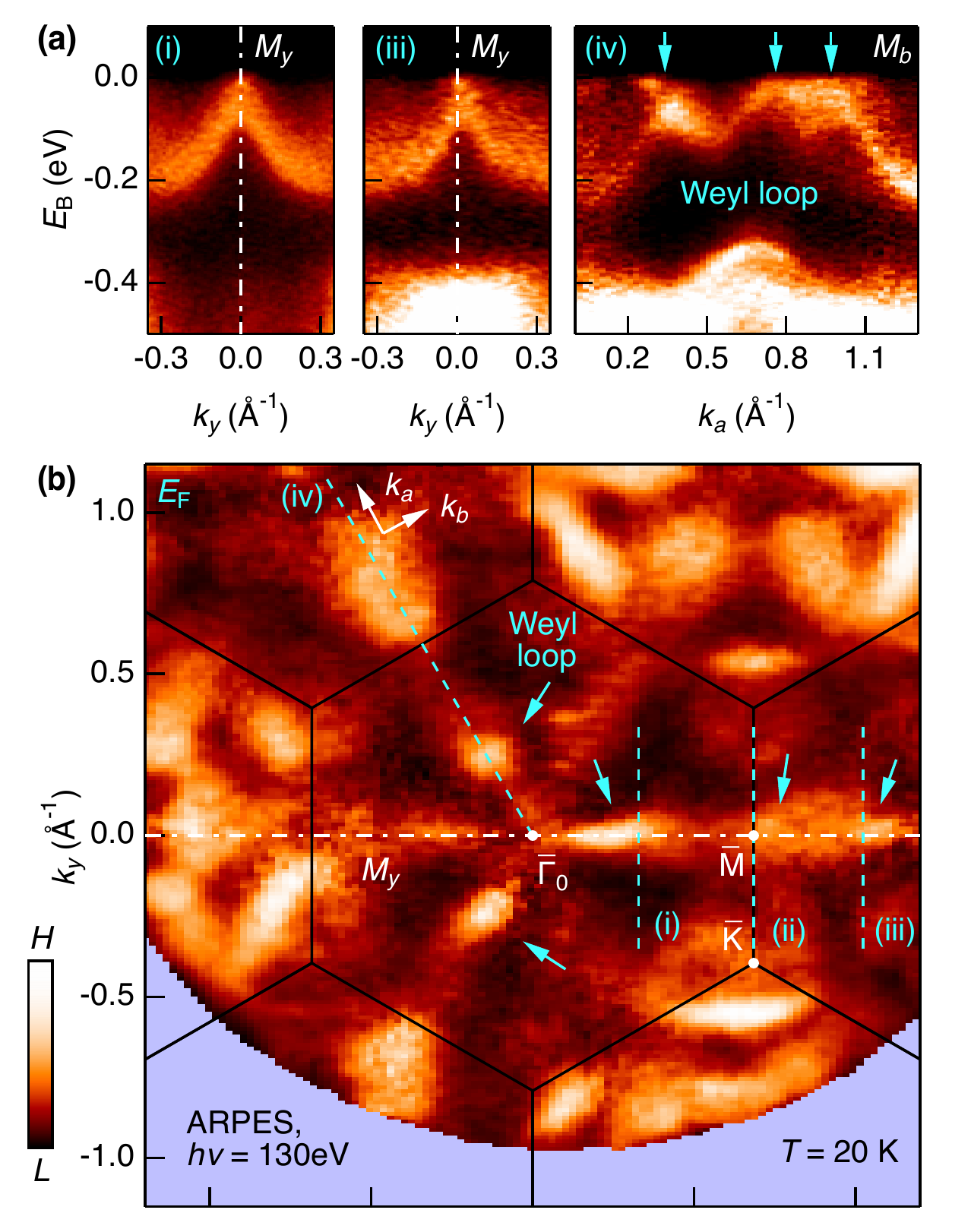}
\caption{\label{WeylLoop_SI_130eV} {\bf Systematics, Weyl loop.} \cpana\ Multiple cone-like dispersions near the Fermi level in energy-momentum ARPES spectra of \s, acquired at $T = 20$ K and photon energy $h\nu = 130$ eV. Cut (i), (iii): transverse to the mirror plane. Cut (iv): contained within the mirror plane. Cut (iv) leftmost arrow: signatures of a hybridization gap with energy scale $\sim 10$ meV, consistent with the moderate SOC in \s\ and suggestive of an SOC-gapped Weyl loop. Cut (ii): see main text. \cpanb\ ARPES Fermi surface exhibiting multiple dot features (cyan arrows) on the mirror planes (corresponding to $\bar{\Gamma}-\bar{M}$). The dots match up with the cone dispersions in (\pana).}
\end{figure}

\clearpage
\begin{figure}
\centering
\includegraphics[width=15cm,trim={0in 0in 0in 0in},clip]{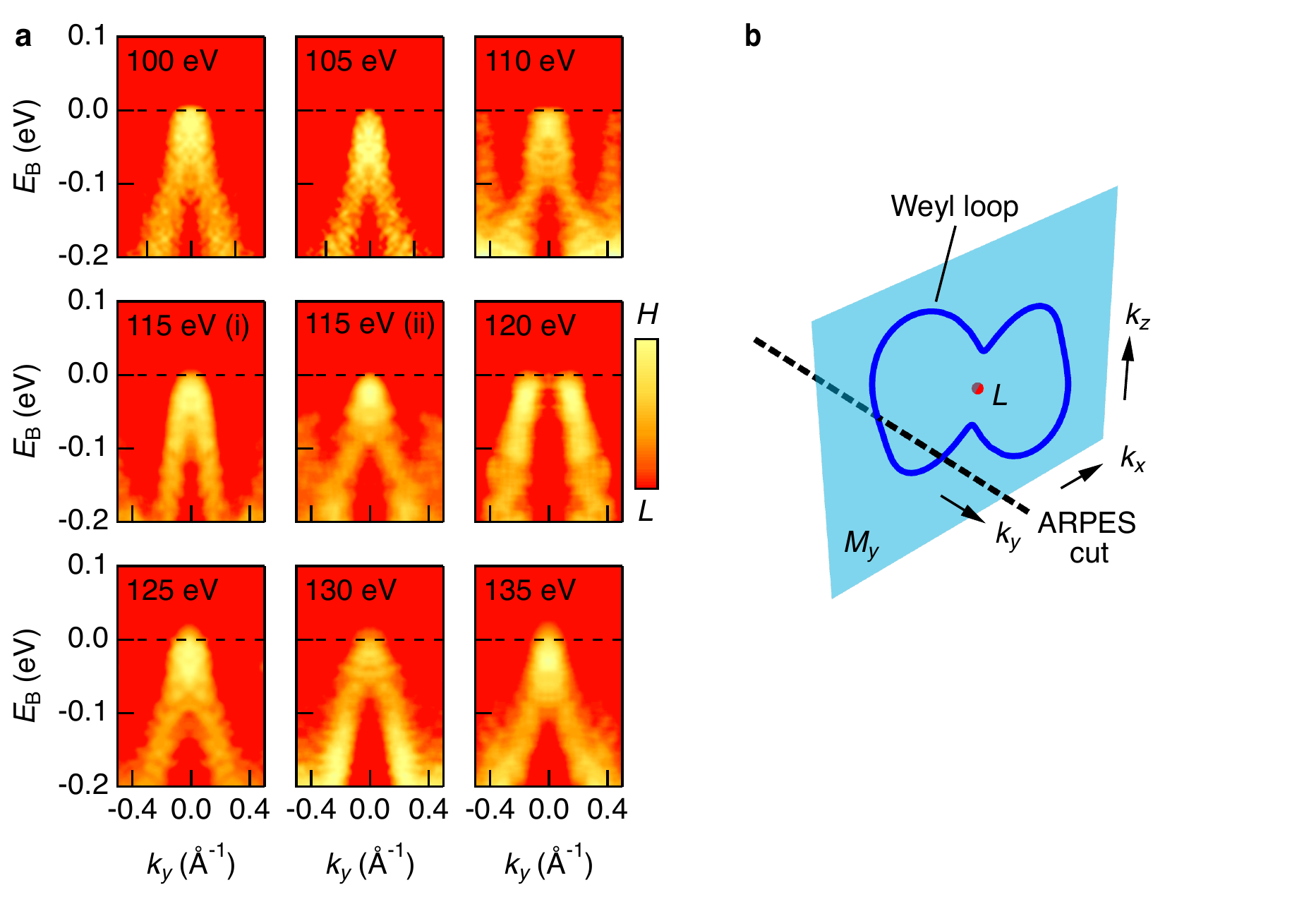}
\caption{\label{WeylLoop_SI} {\bf Systematics, Weyl loop.} \cpana\ Additional source ARPES energy-momentum spectra used to extract the Weyl loop trajectory in Fig. 1f. At each value of $h\nu$ from 100 eV to 135 eV, we examined energy-momentum cuts upon sweeping in $k_x$, collecting all spectra exhibiting cone dispersions. \cpanb\ Converting $h\nu$ to $k_z$, we then assembled the resulting $(k_x,k_z)$ coordinates to obtain the Weyl loop trajectory. Spectra symmetrized around $k_y = 0 {\rm\ \AA}^{-1}$.}
\end{figure}

\clearpage
\begin{figure}
\centering
\includegraphics[width=8cm,trim={0in 0in 0in 0in},clip]{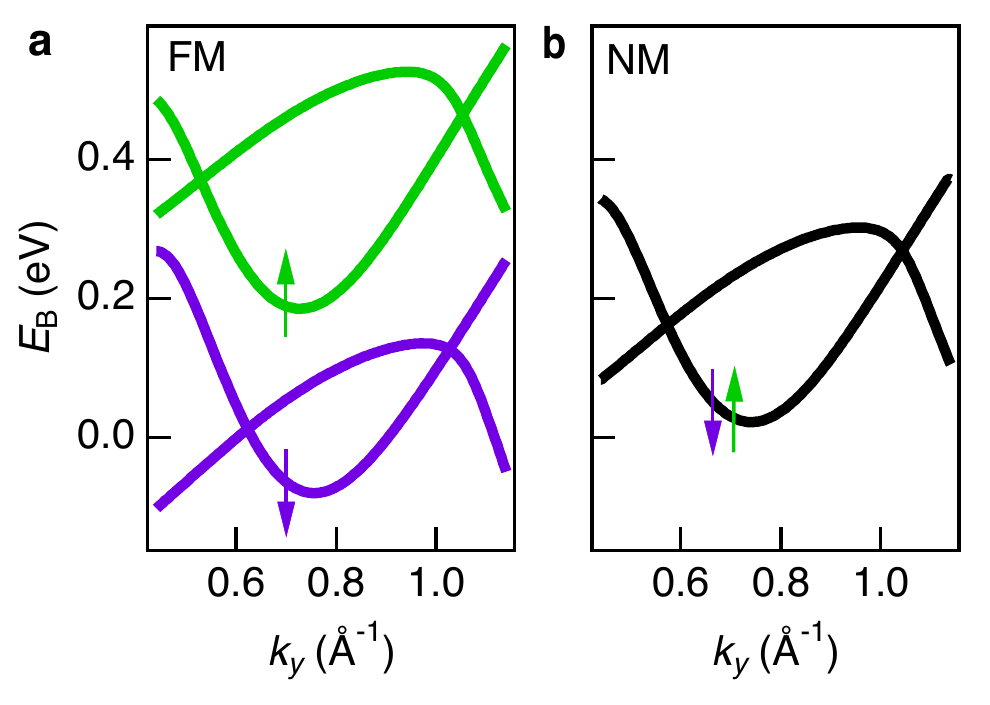}
\caption{\label{DFT_SI} {\bf Weyl to Dirac loop collapse.} \cpana\ \Ab\ calculation in the ferromagnetic state, in the absence of SOC. The Weyl loop near \f\ is formed from two majority spin bands with a partner Weyl loop of the opposite spin $\sim 0.4$ eV above \f. \cpanb\ \Ab\ calculation in the non-magnetic state, in the absence of SOC. The two opposite-spin Weyl loops collapse into a single Dirac loop, characterized by a spinless band crossing along a loop.}
\end{figure}

\clearpage
\begin{figure}
\centering
\includegraphics[width=10cm,trim={0in 0in 0in 0in},clip]{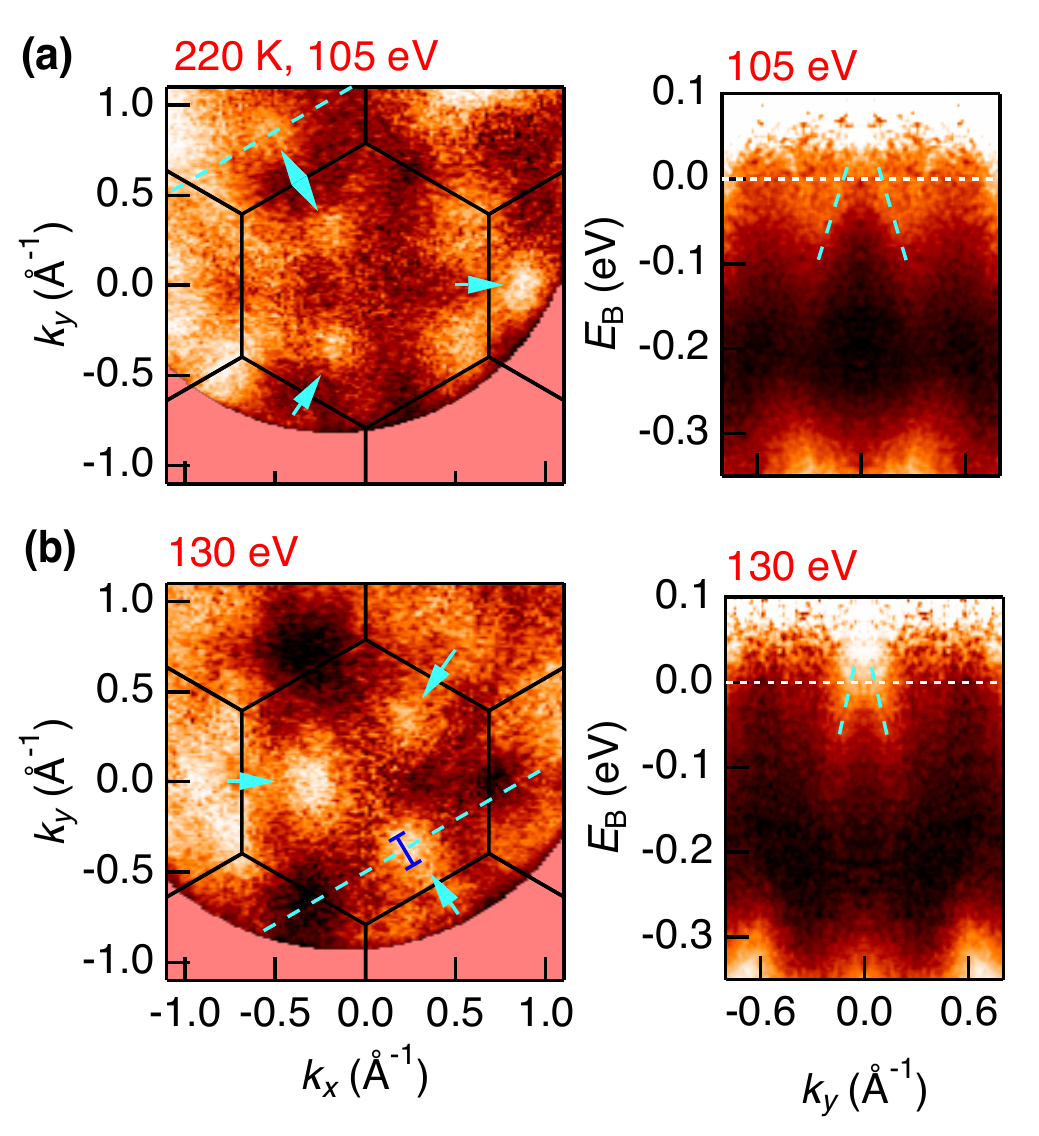}
\caption{\label{DiracLoop_SI} {\bf Systematics, Dirac loop.} \cpana, \cpanb\ Additional source ARPES energy-momentum spectra used to extract the Dirac loop trajectory in Fig. 3e, acquired at $h\nu$ as indicated.}
\end{figure}

\clearpage
\begin{figure}
\centering
\includegraphics[width=17cm,trim={0in 0in 0in 0in},clip]{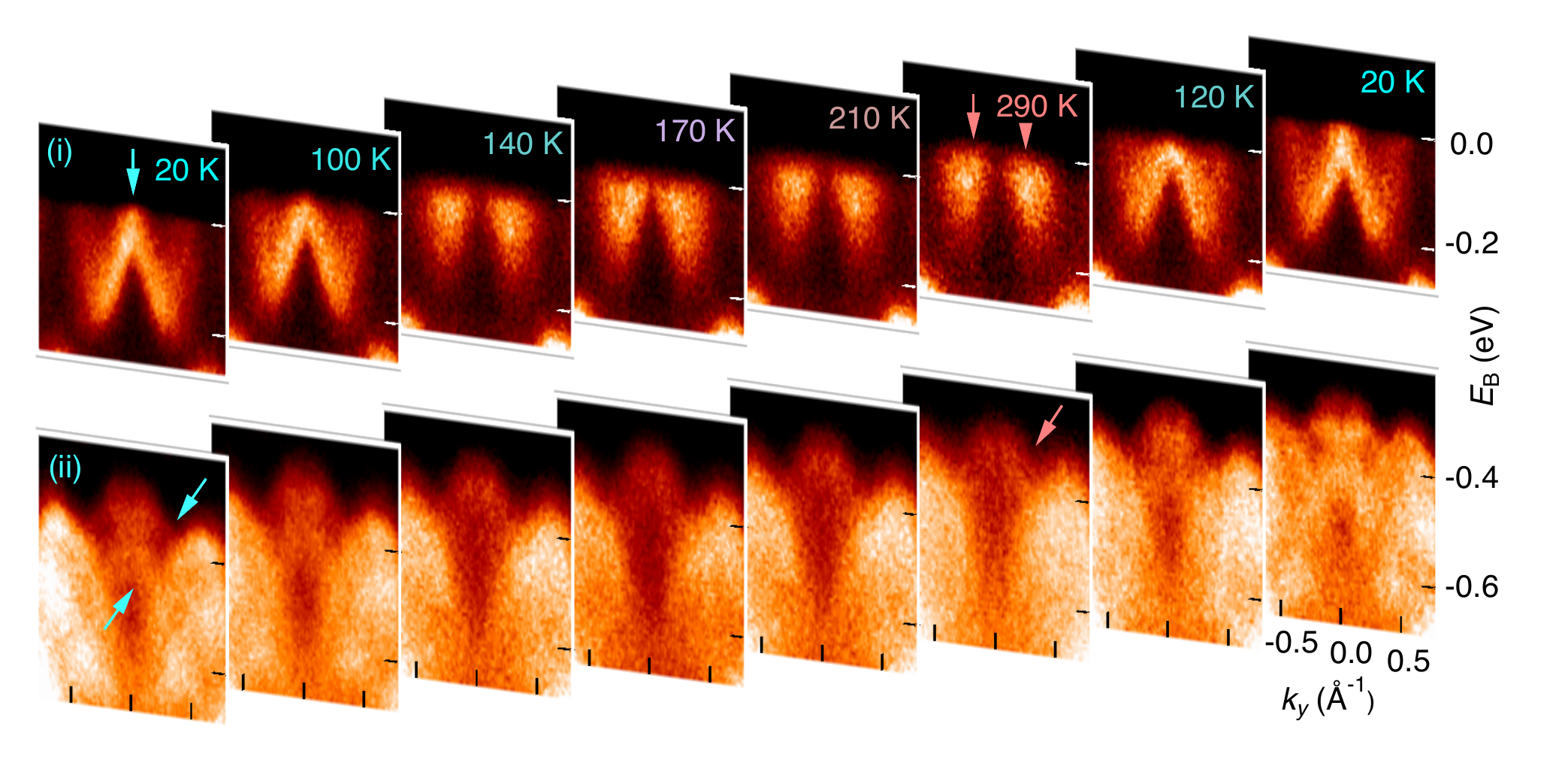}
\caption{\label{tempdep_SI} {\bf Systematics, temperature dependence.} Evolution of the Weyl loop and deep valence bands under temperature cycling, 20 K $\rightarrow$ 290 K $\rightarrow$ 20 K.}
\end{figure}

\clearpage
\begin{figure}
\centering
\includegraphics[width=16cm,trim={0in 0in 0in 0in},clip]{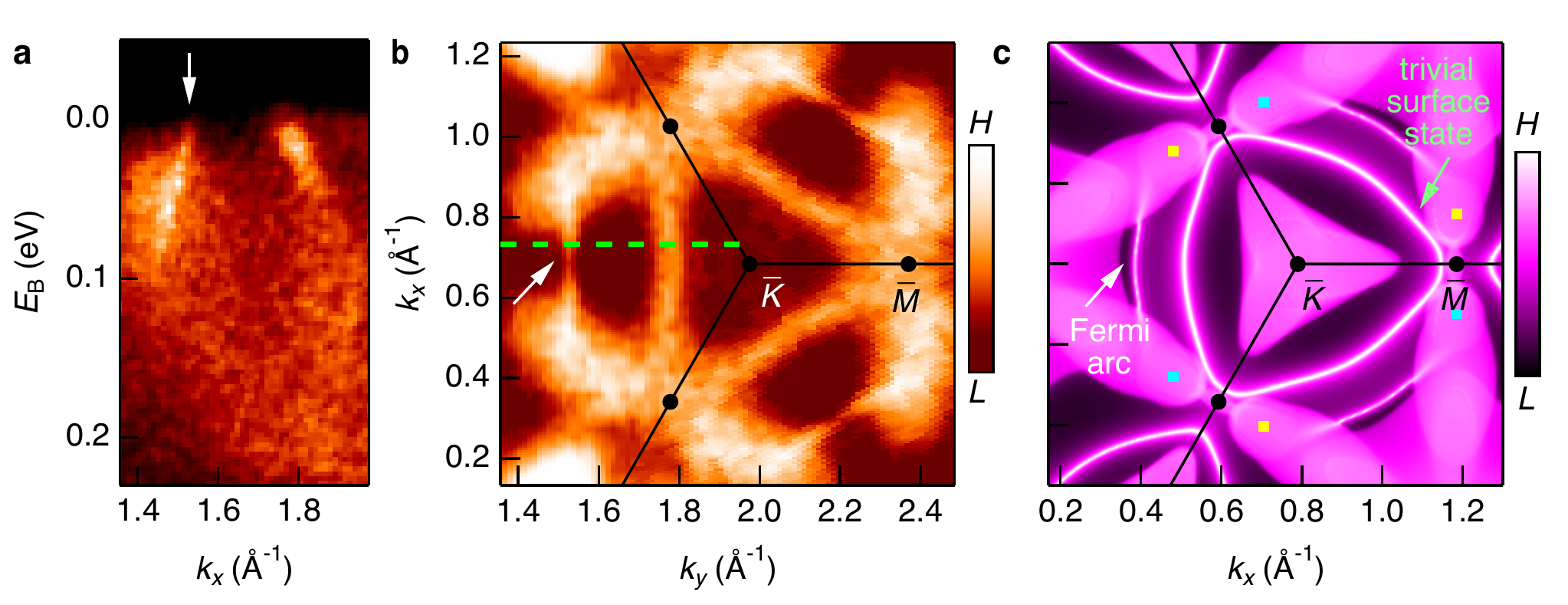}
\caption{\label{arc_SI} {\bf Candidate topological Fermi arc.} \cpana\ Energy-momentum ARPES spectrum and \cpanb\ ARPES Fermi surface in the vicinity of the $\bar{K}$ point, acquired at $h\nu = 130$ eV, linear horizontal light polarization, $\bar{\Gamma}-\bar{K}$ APRES slit alignment, second through fourth surface Brillouin zones, $T = 20$ K. White arrows indicate a candidate topological Fermi arc. \cpanc\ \Ab\ calculation of the surface density of states at the Fermi energy, Sn termination, with SOC. Weyl point predictions marked by the yellow (positive chirality) and cyan (negative chirality) squares, connected by a Fermi arc and near a closed triangular trivial surface state.}
\end{figure}

\clearpage
\begin{figure}
\centering
\includegraphics[width=7cm,trim={0in 0in 0in 0in},clip]{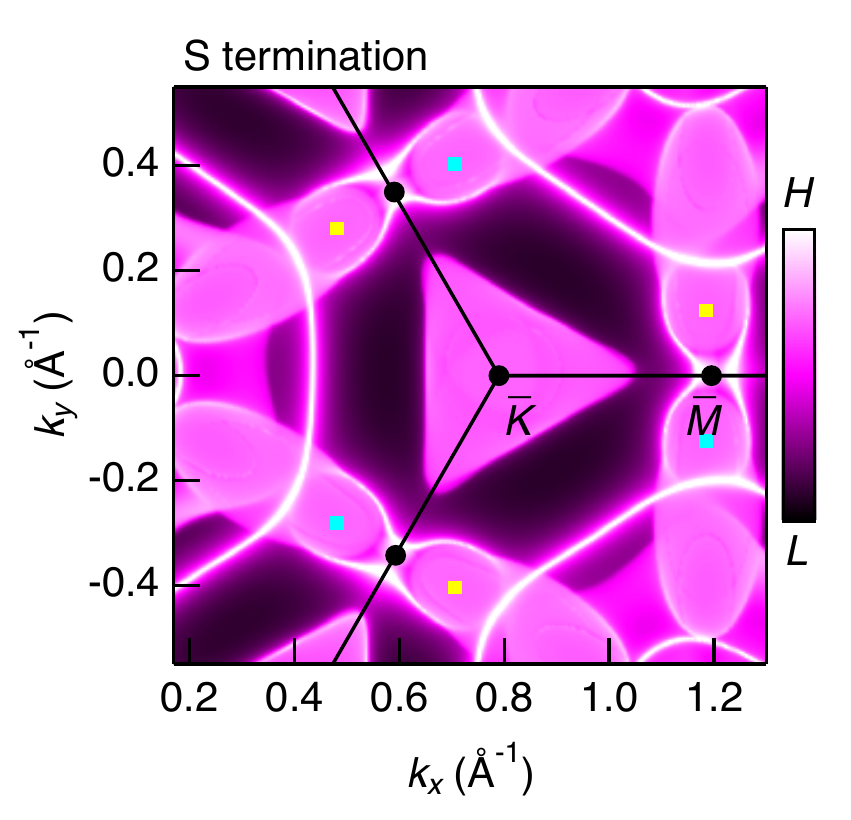}
\caption{\label{Stermination} {\bf Sulfur termination calculation.} \Ab\ calculation of the surface density of states at the Fermi energy, S termination, including spin-orbit coupling, analogous to Fig. \ref{arc_SI}\panc. We observe that our photoemission spectra are best captured by the Sn termination calculation.}
\end{figure}

\begin{figure}
\centering
\includegraphics[width=5cm,trim={0in 0in 0in 0in},clip]{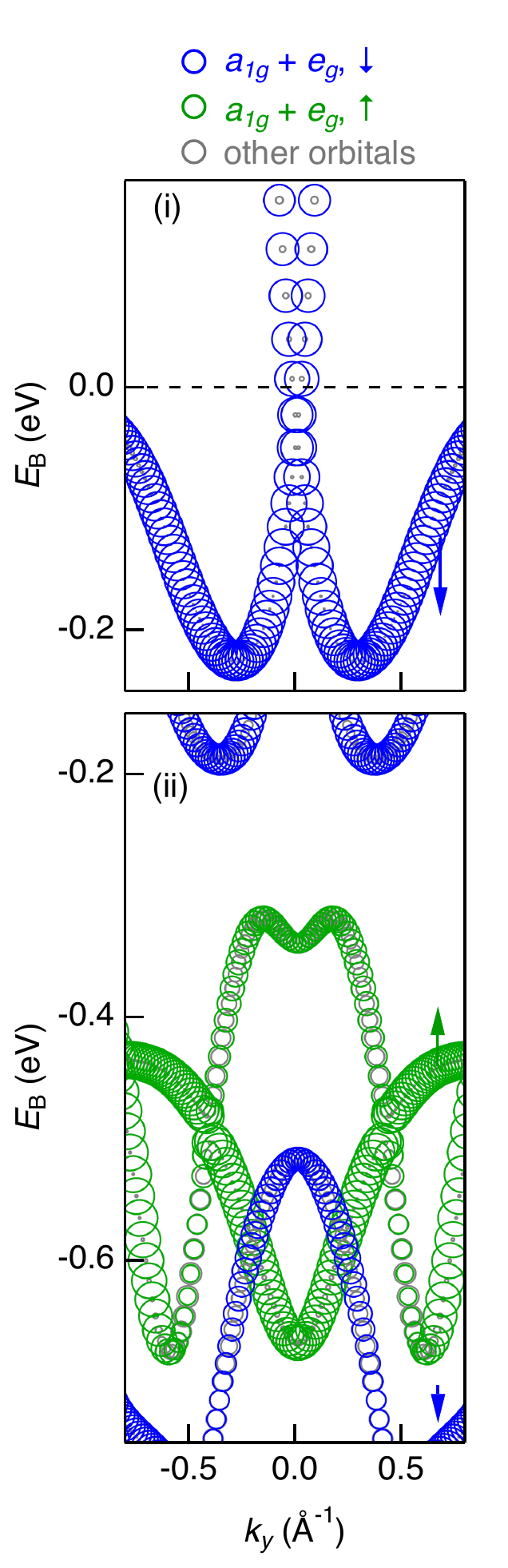}
\caption{\label{ref_orbital} {\bf Orbital decomposition of the Weyl loop and deep valence bands.} Composition on Cut (i) and Cut (ii), ignoring SOC, under FM order (see Fig. 2). The circles mark the fraction of orbital weight originating from the Co $3d$ $a_{1g}$ and $e_g$ manifolds for up and down spin (green \& blue) and all other orbitals (gray). The $a_{1g}$ and $e_g$ manifolds of the $D_{3d}$ point group include $d_{z^2}$, $d_{x^2-y^2}$ and $d_{xy}$, which are all heavily hybridized in the present case \cite{Co3Sn2S2_oxidation_Claudia}. The Weyl loops and deep bands exhibit similar orbital composition.}
\end{figure}